\documentclass[conference]{IEEEtran}

\usepackage{subcaption}
\usepackage{booktabs}
\usepackage{amsmath}
\usepackage{amsfonts}
\usepackage{listings}
\usepackage{color}
\usepackage{xspace}
\usepackage{comment}
\usepackage{graphicx} 
\usepackage{wrapfig}
\usepackage{multirow}
\usepackage{bm}
\usepackage{microtype}
\usepackage{enumitem}
\usepackage{flushend}

\setlist{leftmargin=5.5mm}

\definecolor{listinggreen}{rgb}{0,0.6,0}
\definecolor{listinggray}{rgb}{0.5,0.5,0.5}
\definecolor{listingmauve}{rgb}{0.58,0,0.82}
\definecolor{listingkeywordcolor}{rgb}{1.0,0.4,0.0}
\definecolor{listinglightgray}{rgb}{0.8863,0.8863,0.8863}

\newcommand\yes{\color{listinggreen}{\textbf{\scriptsize Yes}}}
\newcommand\no{\color{listingkeywordcolor}{\textbf{\scriptsize No}}}
\newcommand\limited{\color{listingkeywordcolor}{\textbf{\tiny Limited}}}

\newcommand\COMMENT[1]{}
\newcommand\framework{\textsc{Tiramisu}\xspace}
\newcommand\processor{space\xspace}
\newcommand\Processor{Space\xspace}

\newcommand\Layerone{Abstract Algorithm\xspace}

\newcommand\Layertwo{Computation Management\xspace}
\newcommand\layerthree{data management\xspace}
\newcommand\Layerthree{Data Management\xspace}

\newcommand\Layerfour{Communication Management\xspace}

\newcommand\codeone{(a)\xspace}
\newcommand\codetwo{(b)\xspace}
\newcommand\codethree{(c)\xspace}

\newcommand{\HIDE}[1]{}

\newcommand{\polyc}[1]{
                      {\footnotesize
                       \newline
                       \centerline{\color{listingmauve}{#1}}
                      }
                     }
\newcommand{\poly}[1]{
                      {\footnotesize     
                       \color{listingmauve}{#1}
                      }
                     }

\usepackage[bookmarks=false]{hyperref}
\begin{document}

\title{\framework{}: A Polyhedral Compiler for Expressing Fast and Portable Code}

\author{\IEEEauthorblockN{Riyadh Baghdadi}
\IEEEauthorblockA{\textit{MIT, USA} \\
baghdadi@mit.edu}
\and
\IEEEauthorblockN{Jessica Ray}
\IEEEauthorblockA{\textit{MIT} \\
jray@csail.mit.edu}
\and
\IEEEauthorblockN{Malek Ben Romdhane}
\IEEEauthorblockA{\textit{MIT} \\
malek@mit.edu}
\and
\IEEEauthorblockN{Emanuele Del Sozzo}
\IEEEauthorblockA{\textit{Politecnico di Milano} \\
emanuele.delsozzo@polimi.it}
\and
\IEEEauthorblockN{Abdurrahman Akkas}
\IEEEauthorblockA{\textit{MIT} \\
akkas@mit.edu}
\and
\IEEEauthorblockN{Yunming Zhang}
\IEEEauthorblockA{\textit{MIT} \\
yunming@mit.edu}
\and
\IEEEauthorblockN{Patricia Suriana}
\IEEEauthorblockA{\textit{Google} \\
psuriana@google.com}
\and
\IEEEauthorblockN{Shoaib Kamil}
\IEEEauthorblockA{\textit{Adobe} \\
kamil@adobe.com}
\and
\IEEEauthorblockN{Saman Amarasinghe}
\IEEEauthorblockA{\textit{MIT} \\
saman@mit.edu}
}

\maketitle

\begin{abstract}
This paper introduces \framework{}, a polyhedral framework designed to generate high performance code for multiple platforms including multicores, GPUs, and distributed machines. \framework{} introduces a scheduling language with novel commands to explicitly manage the complexities that arise when targeting these systems.
The framework is designed for the areas of image processing, stencils, linear algebra and deep learning.
\framework{} has two main features: it relies on a flexible representation based on the polyhedral model and it has a rich scheduling language allowing fine-grained control of optimizations. \framework{} uses a four-level intermediate representation that allows full separation between the algorithms, loop transformations, data layouts, and communication.  This separation simplifies targeting multiple hardware architectures with the same algorithm.
We evaluate \framework{} by writing a set of image processing, deep learning, and linear algebra benchmarks and compare them with state-of-the-art compilers and  hand-tuned libraries.  We show that \framework{} matches or outperforms existing compilers and libraries on different hardware architectures, including multicore CPUs, GPUs, and distributed machines.
\end{abstract}

\begin{IEEEkeywords}
Code Optimization, Code Generation, Polyhedral Model, Deep Learning, Tensors, GPUs, Distributed Systems
\end{IEEEkeywords}

\section{Introduction}
\label{sec:intro}

Generating efficient code for high performance systems is becoming more and more difficult as these architectures are increasing in complexity and diversity.
Obtaining the best performance requires complex code and data layout transformations, management of complex memory hierarchies, and efficient data communication and synchronization.

For example, consider generalized matrix multiplication (\texttt{gemm}), which computes $C = \alpha AB + \beta C$ and is a
building block of numerous algorithms, including simulations and convolutional neural networks.  Highly-tuned implementations
require fusing the multiplication and addition loops, as well as applying two-level tiling, vectorization, loop unrolling, array packing~\cite{Goto:2008:AHM:1356052.1356053},
register blocking, and data prefetching.  Furthermore, tuned implementations separate partial tiles from full tiles, since partial tiles cannot fully benefit from the same optimizations.
High performance GPU implementations require even more optimizations, including coalescing memory accesses, managing data movement between global, shared, and register memory, and inserting synchronization primitives.  
Automatically generating such complex code is still beyond the capabilities of state-of-the-art compilers.
The importance of kernels such as \texttt{gemm} motivates vendors to release immensely complex hand-optimized libraries for these kernels.  However, for most users, obtaining this level of performance for their own code is challenging, since the effort required to explore the space of possible implementations is intractable when hand-coding complicated code transformations.

\begin{figure}[t]
\centering
  \begin{minipage}{0.22\textwidth}
    \includegraphics[width=\columnwidth,bb=7 3 476 356]{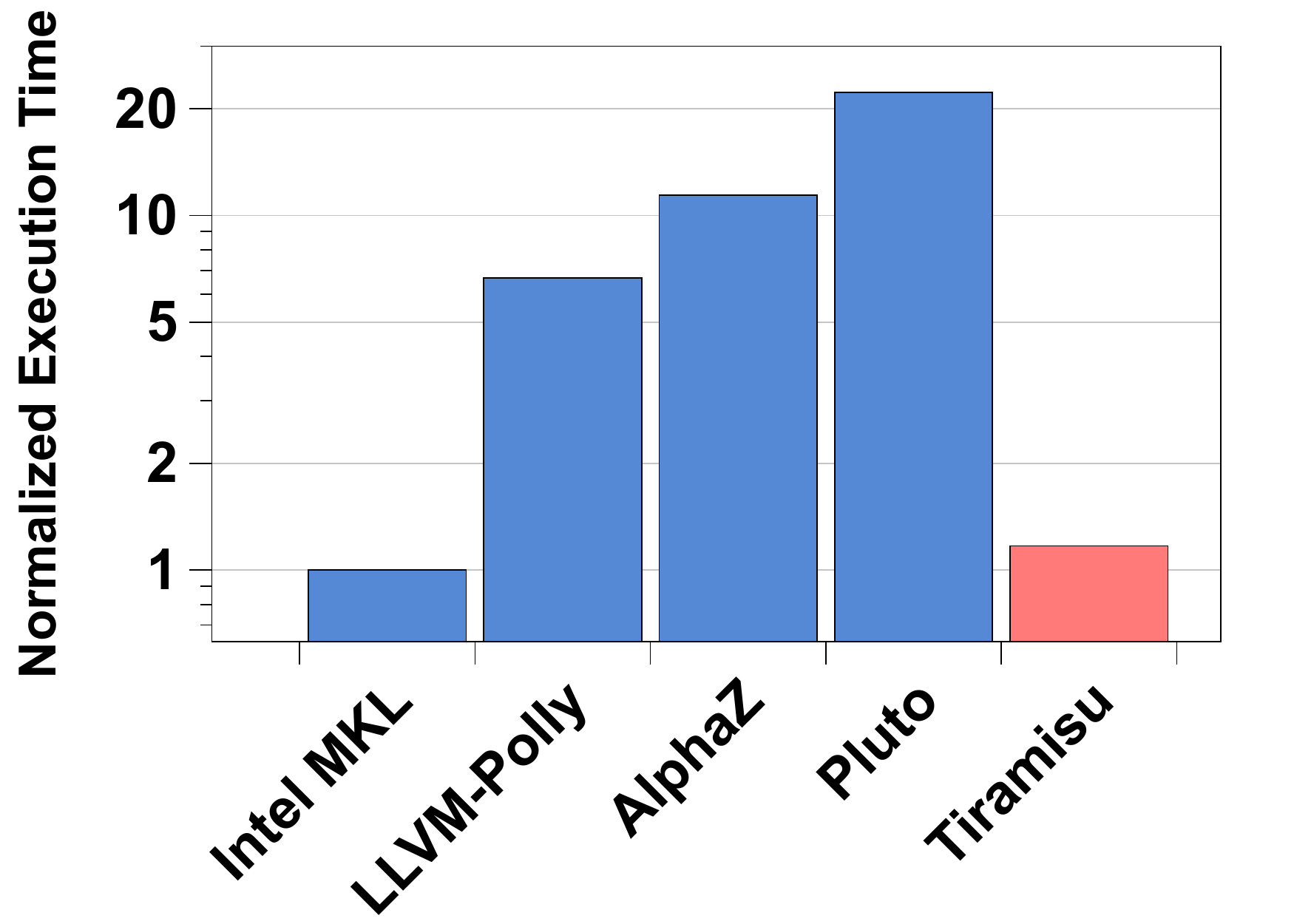}
  \end{minipage}
  \begin{minipage}{0.22\textwidth}
    \includegraphics[width=\columnwidth,bb=7 3 476 356]{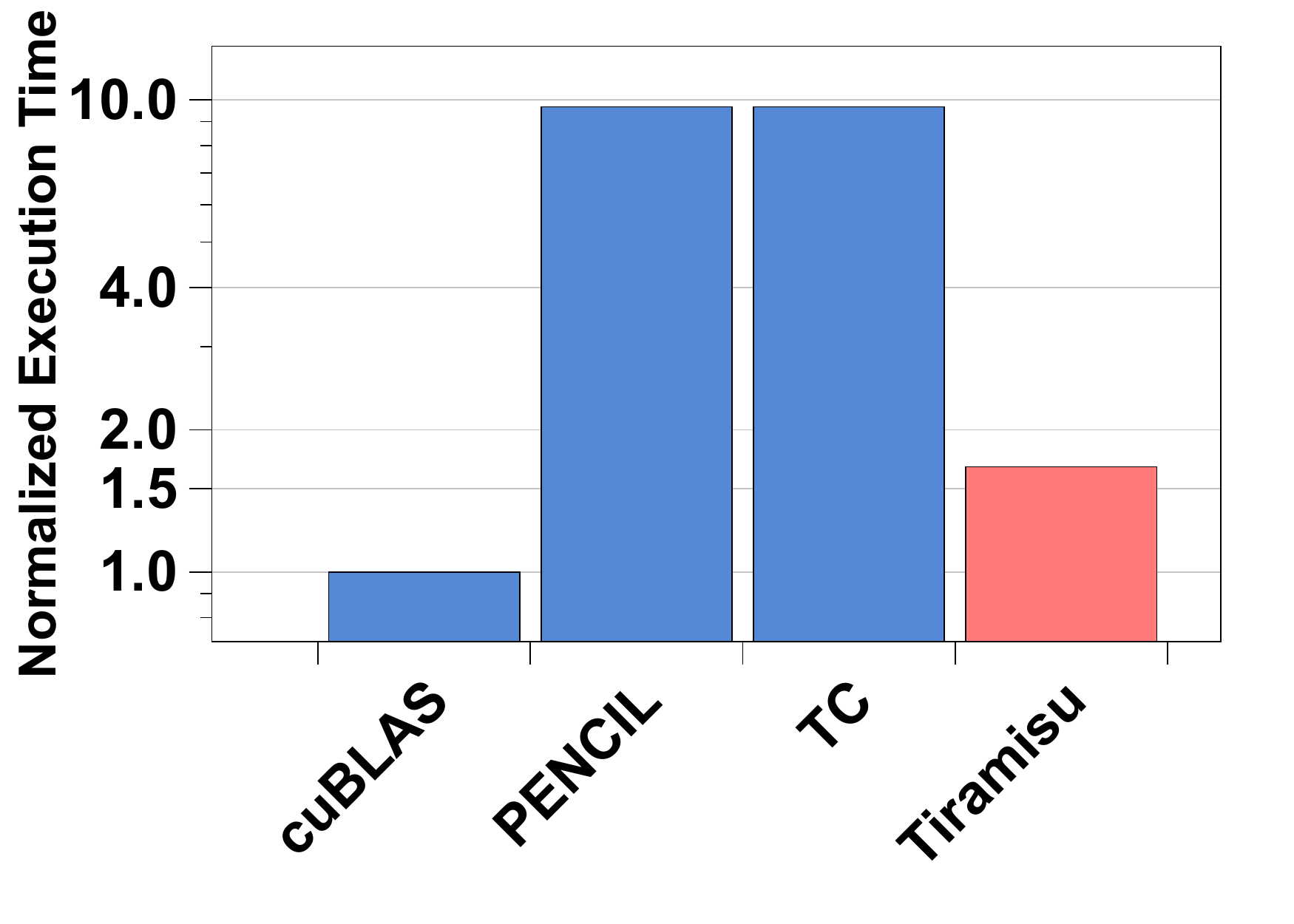}
  \end{minipage}
  \caption{Normalized execution times of code generated for \texttt{sgemm} on CPU (left) and GPU (right).}
  \label{gemm-cpu-gpu}
  \vspace{-0.25cm}
\end{figure}

Previous work using the polyhedral model has shown success in implementing complex iteration space transformations~\cite{wolf1991loop,bondhugula_practical_2008,trifunovic_graphite_2010,polly,Vasilache2018TensorCF,pouchet.11.popl}, data locality optimizations~\cite{Iri88,tobias_hexagonal_cgo13}, and memory management optimizations~\cite{feautrier_array_1988,thies_unified_2001,lefebvre_automatic_1998,Qui00,Darte_contraction_2005}.
Although polyhedral compilers can represent these program and data transformations, they still do not successfully select transformations that result in the best performance. Currently, these compilers do not match the performance of hand-optimized kernels for algorithms such as \texttt{gemm}. The blue bars in Figure~\ref{gemm-cpu-gpu}
show the performance of state-of-the-art polyhedral compilers for \texttt{gemm} compared to the Intel MKL~\cite{mkl} and Nvidia cuBLAS~\cite{cublas} libraries.
Fully-automatic polyhedral compilers such as Polly~\cite{polly} and Pluto~\cite{bondhugula_practical_2008} improve productivity, but do not obtain the desired level of performance since their search techniques consider only a subset of the necessary optimizations and rely on less accurate machine models, leading the compiler to make suboptimal decisions.
Other polyhedral frameworks, such as AlphaZ~\cite{yuki2012alphaz} and CHiLL~\cite{chill}, eschew full automation and instead expose a \textit{scheduling language} that enables users to productively explore the space of possible transformations.
While these frameworks achieve better performance, their scheduling languages are not designed to target distributed systems. For example, they do not allow the user to partition computations, send data across nodes, or insert required synchronization.

In this paper, we introduce \framework{}~\footnote{\url{http://tiramisu-compiler.org/}}, a polyhedral compiler with a scheduling language featuring \emph{novel commands for targeting multiple high performance architectures}.
\framework{} is well-suited for implementing data parallel algorithms (loop nests manipulating arrays).
It takes a high level representation of the program (a pure algorithm and a set of scheduling commands), applies the necessary code transformations, and generates highly-optimized code for the target architecture.
In addition to scheduling commands for loop and data-layout transformations, the \framework{} scheduling language introduces novel commands for explicit communication and synchronization, and for mapping buffers to different memory hierarchies.
In order to simplify the implementation of the scheduling language, \framework{} explicitly divides the intermediate representation into four layers designed to hide the complexity and large variety of execution platforms by separating the architecture-independent algorithm from code transformations, data layout, and communication.
\framework{} targets multicore CPUs, CUDA GPUs, distributed architectures, and FPGA.  This paper  presents the first three backends while Del Sozzo et al.~\cite{8445108} describe an FPGA backend.

The use of a scheduling language has been shown effective for generating efficient code by multiple compilers including CHiLL, AlphaZ, and Halide~\cite{halide_12,DBLP:conf/pldi/Ragan-KelleyBAPDA13}.
In comparison with Halide in particular, not only does \framework{} introduce novel scheduling extensions, \framework{} fundamentally differs in that it relies on the expressive polyhedral representation instead of the interval-based representation used by Halide.  This allows \framework{} to naturally express non-rectangular iteration spaces, to support programs with cyclic data-flow graphs, and to apply any affine transformation (including iteration space skewing), all of which are not naturally expressible in Halide.

This paper makes the following contributions:

\begin{itemize}
  \item We introduce a polyhedral compiler with a scheduling language that features \emph{novel commands for controlling data communication, synchronization, and for mapping to different memory hierarchies}.  These extensions enable targeting multiple high-performance architectures including multicore CPUs, GPUs, and distributed machines.

  \item We explicitly divide the intermediate representation into four layers to simplify the implementation of the scheduling language.  The four-layer IR separates the algorithm from code transformations and data-layout transformations, allowing for portability and simplifying the composition of architecture-specific lowering transformations.

  \item We evaluate \framework{} on a set of deep learning and linear algebra kernels and show that \framework{} can generate efficient code that outperforms Intel MKL by up to $2.3\times$.
  We also evaluate \framework{} on a set of image processing benchmarks and show that \framework{} matches or outperforms state-of-the-art compilers on different hardware architectures, including multicore CPUs, GPUs, and distributed machines.
\end{itemize}


\section{Related Work\label{related}}

\paragraph{Polyhedral compilers with automatic scheduling}

Polyhedral compilers such as PENCIL~\cite{pencil,pencil_paper}, Pluto~\cite{bondhugula_practical_2008}, Polly~\cite{polly}, Tensor Comprehensions~\cite{Vasilache2018TensorCF}, and PolyMage~\cite{Mullapudi:2015:PAO:2786763.2694364} are fully automatic.  Some of them are designed for specific domains (such as Tensor Comprehensions and PolyMage), while Pluto, PENCIL, and Polly are more general.
While fully automatic compilers provide productivity, they may not always obtain the best performance.  This suboptimal performance is due to several reasons: first, these compilers do not implement some key optimizations such as array packing~\cite{Goto:2008:AHM:1356052.1356053}, register blocking, data prefetching, and asynchronous communication (which are all supported by \framework{}); second, they do not have a precise cost-model to decide which optimizations are profitable.  For example, the Pluto~\cite{bondhugula_practical_2008} automatic scheduling algorithm (used in Pluto, PENCIL and Polly) tries to minimize the distance between producer and consumer statements while maximizing outermost parallelism, but it does not consider data layout, redundant computations, or the complexity of the control of the generated code.
Instead of fully automatic scheduling, \framework relies on a set of scheduling commands, giving the user full control over scheduling.


Polyhedral frameworks proposed by Amarasinghe et al.~\cite{Amarasinghe:1993:COC:173262.155102} and Bondhugula et al.~\cite{6877466} address the problem of automatic code generation for distributed systems. Instead of being fully automatic, \framework{} relies on the user to provide scheduling commands to control choices in the generated code (synchronous/asynchronous communication, the granularity of communication, buffer sizes, when to send and receive, cost of communication versus re-computation, etc.).


\paragraph{Polyhedral compilers with a scheduling language}
AlphaZ~\cite{yuki2012alphaz}, CHiLL~\cite{chill,Hall2010} and URUK~\cite{Girbal2006} are polyhedral frameworks developed to allow users to express high-level transformations using scheduling commands. 
Since these frameworks are polyhedral, they can express any affine transformation.
However, their scheduling languages do not target distributed architectures.  In contrast, \framework features scheduling commands for partitioning computations (for distributed systems), synchronization and distribution of data across nodes.  The first four columns of Table~\ref{tab:related} compare between \framework{} and three representative polyhedral frameworks.

\begin{table}[tb]
    \scriptsize
    \vspace{0.5cm}
    \setlength\tabcolsep{1pt}
    \begin{tabular}{l|l|l|l|l|l}
        \hline
        
        \textbf{Feature} & \textbf{Tiramisu} & \textbf{AlphaZ} & \textbf{PENCIL} & \textbf{Pluto} & \textbf{Halide} \\\hline

        \textbf{CPU code generation} & \yes & \yes & \yes & \yes  & \yes \\\hline

        \textbf{GPU code generation} & \yes & \no & \yes & \yes  & \yes \\\hline

        \textbf{Distributed CPU code generation} & \yes & \no & \no & \yes  & \yes\\\hline
        
        \textbf{Distributed GPU code generation} & \yes & \no & \no & \no  & \no\\\hline

        \textbf{Support all affine loop transformations} & \yes & \yes & \yes & \yes  & \no\\\hline

        \textbf{Commands for loop transformations} & \yes & \yes & \no & \no & \yes\\\hline

        \textbf{Commands for optimizing data accesses} & \yes & \yes & \no & \no & \yes \\\hline

        \textbf{Commands for communication} & \yes & \no & \no & \no & \no \\\hline

        \textbf{Commands for memory hierarchies} & \yes & \no & \no & \no & \limited\\\hline

        \textbf{Expressing cyclic data-flow graphs} & \yes & \yes & \yes & \yes & \no \\\hline

        \textbf{Non-rectangular iteration spaces} & \yes & \yes & \yes & \yes & \limited\\\hline

        \textbf{Exact dependence analysis} & \yes & \yes & \yes & \yes  & \no\\\hline
        
        \textbf{Compile-time set emptiness check} & \yes & \yes & \yes & \yes  & \no\\\hline

        \textbf{Implement parametric tiling} & \no & \yes & \no & \no  & \yes\\\hline
    \end{tabular}
    \caption{Comparison between different frameworks.}
    \label{tab:related}
\end{table}

\paragraph{Non-polyhedral compilers with a scheduling language}
Halide~\cite{halide_12} is an image processing DSL with a scheduling language that uses intervals to represent iteration spaces instead of the polyhedral model.  This limits the expressiveness of Halide.
For example, unlike \framework{}, Halide cannot naturally represent non-rectangular iteration spaces, and
this is the reason why distributed Halide~\cite{denniston2016distributed} over-approximates the amount of data to communicate (send and receive) when generating distributed code.
This also makes some Halide passes over-approximate non-rectangular iteration spaces, potentially leading to less efficient code (for example, it prevents Halide from performing precise bounds inference for non-rectangular iteration spaces). The use of intervals also prevents Halide from performing many complex affine transformations, such as iteration space skewing.

Halide does not have dependence analysis and thus relies on conservative rules to determine whether a schedule is legal. For example, Halide does not allow the fusion of two loops (using the \texttt{compute\_with} command) if the second loop reads a value produced by the first loop.
While this rule avoids illegal fusion, it prevents fusing many legal cases, which may lead to suboptimal performance.
Halide also assumes the program has an acyclic dataflow graph in order to simplify checking the legality of a schedule. This prevents users from expressing many programs with cyclic dataflow.
It is possible in some cases to work around the above restrictions, but such work-around methods are not general.
\framework{} avoids over-conservative constraints by relying on dependence analysis to check for the correctness of code transformations, enabling more possible schedules.  Table~\ref{tab:related} summarizes the comparison between \framework{} and Halide.

Vocke et al.~\cite{Vocke:2017:EHI:3132652.3106343} extend Halide to target DSPs, and add scheduling commands such as \texttt{store\_in} to specify in which memory hierarchy data should be stored.
TVM~\cite{tvm} is another system that shares many similarities with Halide. It uses a modified form of the Halide IR internally. Since TVM is also a non-polyhedral compiler, the differences between Halide and \framework{} that are due to the use of polyhedral model also apply to TVM.

POET~\cite{Yi:2007ay} is a system that uses an XML-based description of code and transformation behavior to parametrize loop transformations.  It uses syntactic transformations, which are less general than the polyhedral transformations used in \framework.  GraphIt~\cite{Zhang:2018:GHG:3288538.3276491} is another compiler that has a scheduling language but that is mainly designed for the area of graph applications.

\paragraph{Other Compilers}

Delite \cite{chafi_domain-specific_2011} is a generic framework for building DSL compilers.
It exposes several parallel computation patterns that DSLs can use to express parallelism.
NOVA~\cite{Collins:2014:NFL:2627373.2627375} and Lift~\cite{Steuwer:2017:LFD:3049832.3049841} are IRs for DSL compilers.  They are functional languages that rely on a suite of higher-order functions such as map, reduce, and scan to express parallelism.
\framework{} is complementary to these frameworks as \framework{} allows complex affine transformations that are easier to express in the polyhedral model.

\section{The \framework Embedded DSL}

\framework{} is a domain-specific language (DSL) embedded in C++. It provides a C++ API that allows users to write a high level, architecture-independent algorithm and a set of scheduling commands that guide code generation.
Input \framework code can either be written directly by a programmer, or generated by a different DSL compiler.  \framework{} then constructs a high level intermediate representation (IR), applies the user-specified loop and data-layout transformations, and generates optimized backend code that takes advantage of target hardware features (LLVM IR for multicores and distributed machines and LLVM IR + CUDA for GPUs).

\subsection{Scope of \framework{}}
\framework is designed for expressing data parallel algorithms, especially those that operate over dense arrays using loop nests and sequences of statements.  These algorithms are often found in the areas of image processing, deep learning, dense linear algebra, tensor operations and stencil computations.

\begin{figure}[!t]
\begin{lstlisting}[language=C,escapechar=@]
// Declare the iterators i, j and c.
Var i(0, N-2), j(0, M-2), c(0, 3);@\label{fig:example:tiramisu:iterators}@

Computation bx(i, j, c), by(i, j, c);

// Algorithm.
bx(i,j,c) = (in(i,j,c)+in(i,j+1,c)+in(i,j+2,c))/3;@\label{fig:example:tiramisu:computation1}@
by(i,j,c) = (bx(i,j,c)+bx(i+1,j,c)+bx(i+2,j,c))/3@\label{fig:example:tiramisu:computation2}@);
\end{lstlisting}
\caption{\label{fig:algorithm}\texttt{Blur} algorithm without scheduling commands.}
\end{figure}

\subsection{Specifying the Algorithm}
The first part of a \framework{} program specifies the algorithm without specifying loop optimizations (when and where the computations occur), data layout (how data should be stored in memory), or communication.
At this level there is no notion of data location; rather, values are communicated via explicit producer-consumer relationships.

The algorithm is a pure function that has inputs, outputs, and is composed of a sequence of computations.  A computation is used to represent a statement in \framework.  Flow-control around computations is restricted to \texttt{for} loops and conditionals.  \texttt{While} loops, early exits, and \texttt{GOTO}s cannot be expressed.  To declare a computation, the user provides both the iteration domain of the computation and the expression to compute.  

Figure~\ref{fig:algorithm} shows a blur algorithm written in \framework. This algorithm declares two computations, \texttt{bx} and \texttt{by}.
The first computation, \texttt{bx}, computes a horizontal blur of the input, while the second computation, \texttt{by}, computes the final blur \texttt{by} averaging the output of the first stage.
The iterators \texttt{i}, \texttt{j}, and \texttt{c} in line~\ref{fig:example:tiramisu:iterators} define the iteration domain of \texttt{bx} and \texttt{by} (for brevity we ignore boundary conditions).
The algorithm is semantically equivalent to the following code.

\begin{lstlisting}[language=C,escapechar=@,numbers=none]
for (i in 0..N-2)
 for (j in 0..M-2)
  for (c in 0..3)
   bx[i][j][c] =
        (in[i][j][c]+in[i][j+1][c]+in[i][j+2][c])/3
for (i in 0..N-2)
 for (j in 0..M-2)
  for (c in 0..3)
   by[i][j][c] =
        (bx[i][j][c]+bx[i+1][j][c]+bx[i+2][j][c])/3
\end{lstlisting}

\subsection{Scheduling Commands}


\framework provides a set of high-level scheduling commands for common optimizations; Table~\ref{tab:scheduling} shows some examples.  There are four types of scheduling commands:
\begin{itemize}
    \item Commands for loop nest transformations: these commands include common affine transformations such as loop tiling, splitting, shifting, etc.  For example, applying 32$\times$32 loop tiling to a computation \texttt{C} can be done by calling\\ \texttt{C.tile(i,j,32,32,i0,j0,i1,j1)} where \texttt{i} and \texttt{j} are the original loop iterators and \texttt{i0}, \texttt{j0}, \texttt{i1}, and \texttt{j1} are the names of the loop iterators after tiling.

    \item Commands for mapping loop levels to hardware: examples of these include loop parallelization, vectorization, and mapping loop levels to GPU block or thread dimensions. For example, calling \texttt{C.vectorize(j, 4)} splits the \texttt{j} loop by a factor of 4 and maps the inner loop to vector lanes.

    \item Commands for manipulating data: these include (1) allocating arrays; (2) setting array properties including whether the array is stored in host, device, shared, or local memory (GPU); (3) copying data (between levels of memory hierarchies or between nodes);  and (4) setting array accesses. In most cases, users need only to use high level commands for data manipulation. If the high level commands are not expressive enough, the user can use the more expressive low level commands.

    \item Commands for adding synchronization operations: the user can either declare a barrier or use the \texttt{send} and \texttt{receive} functions for point-to-point synchronization.
\end{itemize}

\begin{table}[h!t]
    \scriptsize
    \vspace{0.5cm}
    \setlength\tabcolsep{1pt}
    \begin{tabular}{l|l}
        \hline
        \multicolumn{2}{c}{\textbf{We assume that C and P are computations, b is a buffer}} \\
        \multicolumn{2}{c}{\textbf{i and j are loop iterators}} \\\hline
        \multicolumn{2}{c}{\textbf{Commands for loop nest transformations}} \\\hline
        \textbf{Command} & \textbf{Description} \\\hline
        \begin{tabular}{ll} \texttt{C.tile(}& \texttt{i,j,t1,t2,}\\ & \texttt{i0,j0,i1,j1)}
        \end{tabular} & 
        \begin{tabular}{l}
        Tile the loop levels (i, j) of the computation C\\
        by $t1\times t2$. The names of the new loop levels\\
        are (i0, j0, i1, j1) where i0 is the outermost\\
        loop level and j1 is the innermost.\end{tabular}\\ \hline
        \texttt{C.interchange(i, j)} & Interchange the i and j loop levels of C.\\\hline
        \texttt{C.shift(i, s)} & Loop shifting (shift the loop level i by s\\
        & iterations). \\ \hline
        \texttt{C.split(i, s, i0, i1)} & Split the loop level i by s. (i0, i1) are the new\\
        & loop levels.\\ \hline
        \texttt{P.compute\_at(C, j)} & Compute the computation \emph{P} in the loop nest of \emph{C}\\
        & at loop level j.  This might introduce redundant\\
        & computations.\\
        \hline
        \texttt{C.unroll(i, v)} & Unroll the loop level i by a factor v.\\\hline
        \texttt{C.after(B, i)} & Indicate that C should be ordered after B at the\\
        & loop level i (they have the same order in all\\
        & the loop levels above i).\\
        \hline
        \texttt{C.inline()} & Inline C in all of its consumers.\\ \hline
        \texttt{C.set\_schedule()} & Transform the iteration domain of C using an\\
        & affine  relation (a \emph{map} to transform Layer I to\\
        & II expressed in the ISL syntax).\\\hline
        \multicolumn{2}{c}{\textbf{Commands for mapping loop levels to hardware}} \\\hline
        \texttt{C.parallelize(i)} & Parallelize the i loop level for execution on a\\
        & shared memory system.\\\hline
        \texttt{C.vectorize(i, v)} & Vectorize the loop level i by a vector size v.\\\hline
        \texttt{C.gpu(i0, i1, i2, i3)} & Mark the loop levels i0, i1, i2 and i3 to be\\
        & executed on GPU. (i0, i1) are mapped to block\\
        & IDs and (i2, i3) to thread IDs.\\\hline
        \texttt{C.tile\_gpu(i0,i1,t1,t2)} & Tile the loops i0 and i1 by $t1 \times t2$ and map\\
        & them to GPU.\\\hline
        \texttt{C.distribute(i)} & Parallelize the loop level i for execution on a\\
        & distributed memory system.\\\hline
        \multicolumn{2}{c}{\textbf{High level commands for data manipulation}} \\\hline
        \texttt{\textbf{C.store\_in(b,\{i, j\})}} & Store the result of the computation C(i,j) in b[i,j].\\\hline
        \texttt{\textbf{C.cache\_shared\_at(P,i)}} & Cache (copy) the buffer of C in shared memory.\\
        & Copying from global to shared GPU memory\\
        & will be done at loop level i of the computation P.\\
        & The amount of data to copy,
        the access functions,\\
        & and synchronization are computed automatically.\\\hline
        \texttt{\textbf{C.cache\_local\_at(P, i)}} & Similar to \texttt{cache\_shared\_at} but stores in\\
        & local GPU memory.\\\hline
        \texttt{\textbf{send(d, src, s, q, p)}}
            & Create a send operation. \texttt{d}: vector of iterators\\
            & to represent the iteration domain of the send;\\
            & \texttt{src}: source buffer; \texttt{s}: size;
            \texttt{q}: destination node;\\
            & \texttt{p}: properties (synchronous, 
            asynchronous,\\
            & blocking, ...).
        \\\hline
        \texttt{\textbf{receive(d,dst,s,q,p)}}
            & Create a receive operation. Arguments similar \\
            & to \texttt{send} except \texttt{q}, which is the source node.
        \\\hline
        \multicolumn{2}{c}{\textbf{Low level commands for data manipulation}}\\\hline
        \texttt{\textbf{Buffer b(sizes, type)}} & Declare a buffer (\texttt{sizes}: a vector of dimension\\
        & sizes).\\\hline
        \texttt{\textbf{b.allocate\_at(p, i)}} & Return an operation that allocates b at the loop\\
        & i of p. An operation can be scheduled like\\
        & any computation.
        \\\hline
        \texttt{\textbf{C.buffer()}} & Return the buffer associated to the computation\\
        & C.\\\hline
        \texttt{\textbf{b.set\_size(sizes)}} & Set the size of a buffer. \texttt{sizes}: a vector of\\
        & dimension sizes.\\\hline
        \texttt{\textbf{b.tag\_gpu\_global()}} & Tag buffer to be stored in global GPU memory.\\\hline
        \texttt{\textbf{b.tag\_gpu\_shared()}} & Tag buffer to be stored in shared GPU memory.\\\hline
        \texttt{\textbf{b.tag\_gpu\_local()}} & Tag buffer to be stored in local GPU memory.\\\hline
        \texttt{\textbf{b.tag\_gpu\_constant()}} & Tag buffer to be stored in constant GPU memory.\\\hline
        \texttt{\textbf{C.host\_to\_device()}} & Return an operation that copies C.buffer() from\\
        & host to device.\\\hline
        \texttt{\textbf{C.device\_to\_host()}} & Return an operation that copies C.buffer() from\\
        & device to host.\\\hline
        \texttt{\textbf{copy\_at(p, i, bs, bd)}}
            & Return an operation that copies the buffer \texttt{bs} \\
            & to the buffer \texttt{bd} at the loop i of p.
            Used for\\
            & copies between global, shared and local.
        \\\hline
        \multicolumn{2}{c}{\textbf{Commands for synchronization}} \\\hline
        \texttt{\textbf{barrier\_at(p, i)}}
            & Create a barrier at the loop p of i.
        \\\hline
    \end{tabular}
    \caption{Examples of \framework{} Scheduling Commands}
    \label{tab:scheduling}
\end{table}

Novel commands introduced by \framework are highlighted in \textbf{bold} in Table~\ref{tab:scheduling}.  They include array allocation, copying data between memory hierarchies, sending and receiving data between nodes, and synchronization.  Calls to
\texttt{cache\_shared\_at()}, \texttt{cache\_local\_at()}, \texttt{allocate\_at()}, \texttt{copy\_at()}, \texttt{barrier\_at()} return an operation that can be scheduled like any other computation (an operation in \framework is a special type of computation that does not return any value).
The operations \texttt{cache\_shared\_at()} and \texttt{cache\_local\_at()} can be used to create a cache for a buffer (GPU only).  They automatically compute the amount of data needing to be cached, perform the data copy, and insert any necessary synchronization.

\begin{figure*}[h!t]
    \centering
    \scriptsize
    \setlength\tabcolsep{4pt}
\begin{tabular}{cl|@{}l}
 &
    \ \ \ \ \ \ \ \ \ \ \ \ \ \ \ \ \ \ \ \ \ \ \ \ \ \ \ \ \ \ \ \ \ \ \ 
    \textbf{\framework Scheduling Commands}
 &
    \ \ \ \ \ \ \ \ \ \ \ \ \ \ \ \ \ \ \ \ \ \
    \textbf{Pseudocode Representing Code Generated by \framework}
\\\hline

{\textbf{\normalsize\codeone{}}} &

\begin{lstlisting}[language=C,escapechar=@]
// Scheduling commands for targeting
// a multicore architecture.

// Tiling and parallelization.
Var i0, j0, i1, j1;
by.tile(i, j, 32, 32, i0, j0, i1, j1);
by.parallelize(i0);
bx.compute_at(by, j0);
\end{lstlisting}


& 

\begin{lstlisting}[language=C,escapechar=@]

 @{\color{listingkeywordcolor}{\textbf{Parallel}}}@ for(i0 in 0..floor((N-2)/32))
   for(j0 in 0..floor((M-2)/32))
    bx[32,34,3];
    // Tiling with redundancy
    for(i1 in 0..min((N-2)%32,32)+2)
     for(j1 in 0..min((M-2)%32,32)+2)
     int i = i0*32+i1
     int j = j0*32+j1
     for (c in 0..3)
      bx[i1][j1][c]=@\label{fig:motivating:code1:stmt1}@
        (in[i][j][c] + in[i][j+1][c]
                     + in[i][j+2][c])/3

    for(i1 in 0..min(N-2,32))
     for(j1 in 0..min(M-2,32))
     int i = i0*32+i1
     int j = j0*32+j1
     for (c in 0..3)
      by[i][j][c]=@\label{fig:motivating:code1:stmt2}@
        (bx[i][j][c] + bx[i+1][j][c]
                     + bx[i+2][j][c])/3

\end{lstlisting}
    

\\\hline

{\textbf{\normalsize\codetwo{}}} &

\begin{lstlisting}[language=C,escapechar=@]
// Scheduling commands for targeting GPU.

// Tile i and j and map the resulting dimensions
// to GPU
Var i0, j0, i1, j1;
by.tile_gpu(i, j, 32, 32, i0, j0, i1, j1);
bx.compute_at(by, j0);
bx.cache_shared_at(by, j0);

// Use struct-of-array data layout
// for bx and by.
bx.store_in({c,i,j});
by.store_in({c,i,j});

// Create data copy operations
operation cp1 = in.host_to_device();
operation cp2 = by.device_to_host();

// Specify the order of execution of copies
cp1.before(bx, root);
cp2.after(by, root);
\end{lstlisting}


& 

\begin{lstlisting}[language=C,escapechar=@]

 host_to_device_copy(in_host, in);

 @{\color{listingkeywordcolor}{\textbf{GPUBlock}}}@ for(i0 in 0..floor((N-2)/32))
  @{\color{listingkeywordcolor}{\textbf{GPUBlock}}}@ for(j0 in 0..floor((M-2)/32))
   @{\color{listingkeywordcolor}{\textbf{shared}}}@ bx[3,32,34];
   // Tiling with redundancy
   @{\color{listingkeywordcolor}{\textbf{GPUThread}}}@ for(i1 in 0..min((N-2)%32,32)+2)
    @{\color{listingkeywordcolor}{\textbf{GPUThread}}}@ for(j1 in 0..min((M-2)%32,32)+2)
     int i = i0*32+i1
     int j = j0*32+j1
     for (c in 0..3)
      bx[c][i1][j1]=@\label{fig:motivating:code2:stmt2}@
        (in[i][j][c] + in[i][j+1][c]
                     + in[i][j+2][c])/3

   @{\color{listingkeywordcolor}{\textbf{GPUThread}}}@ for(i1 in 0..min(N-2,32))
    @{\color{listingkeywordcolor}{\textbf{GPUThread}}}@ for(j1 in 0..min(M-2,32))
     int i = i0*32+i1
     int j = j0*32+j1
     for (c in 0..3)
      by[c][i][j]=@\label{fig:motivating:code2:stmt3}@
        (bx[c][i][j] + bx[c][i+1][j]
                     + bx[c][i+2][j])/3

 device_to_host_copy(by, by_host);

\end{lstlisting}
    

\\\hline
{\textbf{\normalsize\codethree{}}} &
\begin{lstlisting}[language=C,escapechar=@]
// Scheduling commands for targeting
// a distributed system

// Declare additional iterators
Var is(1, Nodes), ir(0,Nodes-1), i0, i1;

// Split loop i into loops i0 and i1 and
// parallelize i1
bx.split(i,N/Ranks,i0,i1); bx.parallelize(i1);@\label{line:split1}@
by.split(i,N/Ranks,i0,i1); by.parallelize(i1);@\label{line:split2}@ 

// Communicate the border rows where necessary
send s =
  send({is}, lin(0,0,0), M*2*3, is-1, {ASYNC});@\label{line:send}@
recv r =
  receive({ir}, lin(N,0,0), M*2*3, ir+1,{SYNC},s);@\label{line:recv}@
  
// Order execution
s.before(r,root);
r.before(bx,root)

// Distribute the outermost loops
bx.distribute(i0); by.distribute(i0); @\label{line:comp_dist}@
s.distribute(is); r.distribute(ir); @\label{line:comm_dist}@
\end{lstlisting}

& 


\begin{lstlisting}[language=C,escapechar=@]
 // We assume that in[][][] is initially
 // distributed across nodes.  Each node
 // has a chunk of the original
 // in[][][] that we call lin[][][].

 // Start by exchanging border rows of
 // lin[][][]
 @{\color{listingkeywordcolor}{\textbf{distributed}}}@ for (is in 1..Nodes) @\label{line:distfor}@
   send(lin(0,0,0), M*2*3, is-1,{ASYNC})
 @{\color{listingkeywordcolor}{\textbf{distributed}}}@ for (ir in 0..Nodes-1)
   recv(lin(N,0,0), M*2*3, ir+1, {SYNC})

 @{\color{listingkeywordcolor}{\textbf{distributed}}}@ for (i0 in 0..Nodes)
   @{\color{listingkeywordcolor}{\textbf{parallel}}}@ for (i1 in 0..(N-2)/Nodes)
    int i = i0*((N-2)/Nodes) + i1
    for (j in 0..M-2)
      for (c in 0..3)
        bx[i][j][c] =@\label{fig:motivating:code2:stmt1}@
          (lin[i][j][c] + lin[i][j+1][c]
                        + lin[i][j+2][c])/3

 @{\color{listingkeywordcolor}{\textbf{distributed}}}@ for (i0 in 0..Nodes)
   @{\color{listingkeywordcolor}{\textbf{parallel}}}@ for (i1 in 0..(N-2)/Nodes)
    int i = q*((N-2)/Nodes) + i1
    for (j in 0..M-2)
      for (c in 0..3)
        by[i][j][c] =
            (bx[i][j][c] + bx[i+1][j][c]
                         + bx[i+2][j][c])/3

 // We assume that no gather operation on
 // by[][][] is needed
\end{lstlisting}


\\\hline
\end{tabular}
\caption{Three examples illustrating \framework{} scheduling commands (left) and the corresponding generated code (right). \codeone{} shows scheduling commands for mapping to a multicore architecture; \codetwo{} shows scheduling commands for mapping to GPU; \codethree{} uses commands to map to a distributed CPU machine.}
\label{fig:mainexample}
\end{figure*}

The use of \texttt{allocate\_at()}, \texttt{copy\_at()}, and \texttt{barrier\_at()} allows \framework to automatically compute iteration domains for the data copy, allocation, and synchronization operations.  This is important because it relieves the user from guessing or computing the iteration domain manually, especially when exploring different possible schedules.  For example, consider copying a buffer from global memory to shared memory in a loop nest executing on a GPU.  The size of the area to copy and the iteration domain of the copy operation itself (which is a simple assignment in this case) depends on whether the loop is tiled, the tile size, and  whether any other loop transformation has already been applied.  \framework simplifies this step by automatically computing the iteration domain and the area of data to copy from the schedule.

To illustrate more \framework{} scheduling commands, let us take the \texttt{blur} example again from Figure~\ref{fig:algorithm} and map it for execution on a multicore architecture.
The necessary scheduling commands are shown in Figure~\ref{fig:mainexample}-\codeone{} (left).
The \texttt{tile()} command tiles the computation \texttt{by}.  The \texttt{compute\_at()} command computes the tile of \texttt{bx} that needs to be consumed by \texttt{by} at the loop level \texttt{j0}.  This transformation introduces redundant computations (in this case) and is known as overlapped tiling~\cite{Krishnamoorthy:2007:EAP:1273442.1250761}. The \texttt{parallelize()} command parallelizes the \texttt{i0} loop.

Now let us take the same example but map the two outermost loops of \texttt{bx} and \texttt{by} to GPU.  The necessary scheduling commands are shown in Figure~\ref{fig:mainexample}-\codetwo{} (left).
The \texttt{tile\_gpu()} command tiles the computation \texttt{by} then maps the new loops to GPU block and thread dimensions.  The \texttt{compute\_at()} command computes the tile of \texttt{bx} needed by \texttt{by} at the loop level \texttt{j0} (this introduces redundant computations).
\texttt{cache\_shared\_at()} instructs \framework to store the results of the \texttt{bx} computation in shared memory. Copying from global to shared memory will be done at the loop level \texttt{j0} of \texttt{by}.
The subsequent \texttt{store\_in()} command  specifies the access functions for \texttt{bx} and \texttt{by}.  In this case, it indicates that these computations are stored in a SOA (struct-of-array) data layout (to allow for coalesced accesses).  The final commands create data copy operations (host-to-device and device-to-host) and schedule them.

Suppose we want to run the \texttt{blur} example on a distributed system with a number of multicore CPU nodes equal to \texttt{Nodes}. Figure \ref{fig:mainexample}-\codethree{} (left) shows the scheduling commands to use in this case. We assume that the array \texttt{in[][][]} is initially distributed across nodes such that node \texttt{n} has the chunk of data represented by 
\texttt{in[n*((N-2)/Nodes)..(n+1)*((N-2)/Nodes),*,*]}. In other words, this corresponds to row \texttt{n*(N-2)/Nodes} through row \texttt{(n+1)*((N-2)/Nodes)}. This chunk is stored in the local array \texttt{lin[][][]}. 

\texttt{send()} and \texttt{recv()} define communication for the border regions. Assuming that each node has a chunk of \texttt{in}.  The \texttt{blur} computation for a chunk stored in node \texttt{n} requires the first two rows of data from the chunk stored in node \texttt{n+1}.  These two rows are referred to as the border region.
The \texttt{send()} will send 2 rows ($M\times 2 \times3$ contiguous data elements) from node \texttt{is} to node \texttt{is-1} starting from \texttt{lin(0,0,0)}, which corresponds to the first two rows of the chunk on node \texttt{is}. In response, the \texttt{recv} for node \texttt{ir} will receive 2 rows ($M\times 2 \times3$ contiguous data elements) from node \texttt{ir+1}, which corresponds to \texttt{ir} receiving the first two rows from node \texttt{ir+1}. The receive for node \texttt{ir} places these elements starting at the end of its local chunk by starting at \texttt{lin(N,0,0)}.
Additionally, \texttt{\{ASYNC\}} defines an asynchronous send and \texttt{\{SYNC\}} defines a synchronous receive. Finally, we tag the appropriate loops (the outer loops of \texttt{bx}, \texttt{by}, \texttt{s}, and \texttt{r}), to be distributed (i.e., we tag each iteration to run on a different node).

All other scheduling commands in \framework{} can be composed with \texttt{send}s, \texttt{recv}s, and distributed loops, as long as the composition is semantically correct. 


\section{\label{sec:ir}The \framework{} IR}
The main goal of \framework{}'s multi-layer intermediate representation is to simplify the implementation of scheduling commands by applying them in a specific order.
This section illustrates why, and describes the layers of the \framework{} IR.

\subsection{Rationale for a Multi-layer IR}

In this section we provide examples showing why current intermediate representations are not adequate for \framework{} and why we need a multi-layer IR.

Most current intermediate representations use memory to communicate between program statements.  This creates memory-based dependencies in the program, and forces compilers to choose data layout before deciding on optimizations and mapping to hardware.  Optimizing a program for different hardware architectures usually requires modifying the data layout and eliminating memory-based dependencies since they restrict optimizations~\cite{maydan1992data}.  Thus, any data layout specified before scheduling must be undone to allow more freedom for scheduling, and the code must be adapted to use the data-layout best-suited for the target hardware.
Applying these data-layout transformations and the elimination of memory-based dependencies is challenging~\cite{gupta1997privatization,autoPrivatPeng,li_array_1992,feautrier_array_1988,midkiff_automatic_2012,maydan_array-data_1993,lefebvre_automatic_1998,Qui00,Darte_contraction_2005}.

Another example that demonstrates the complexity of code generation is mapping buffers to shared and local memory on GPU.  The amount of data that needs to be copied to shared memory and when to perform synchronization both depend on how the code is optimized (for example,  whether the code has two-level tiling or not).  The same applies to deciding the amount of data to send or receive when generating distributed code.   Therefore, buffer mapping to memory hierarchies, communication management, and synchronization should not occur before scheduling.

\framework{} addresses these complexities in code generation by using a multi-layer IR that fully separates the architecture-independent algorithm from loop transformations, data layout and communication.
The \emph{first layer} representation describes the pure algorithm using producer-consumer relationships without memory locations.
The \emph{second layer} specifies the order of computation, along with which processor computes each value; this layer is suitable for performing a vast number of optimizations without dealing with concrete memory layouts.
The \emph{third layer} specifies where to store intermediate data before they are consumed.  The \emph{fourth layer} adds all the necessary communication and synchronization operations.

The separation of layers defines a specific order for applying optimizations and ensures that compiler passes in a given layer need not to worry about modifying or undoing a decision made in an earlier layer.  For example, the phase that specifies the order of computations and where they occur can safely assume that no data-layout transformations are required.
This simple assumption allows \framework{} to avoid the need to rely on a large body of research that focuses on data-layout transformations to allow scheduling~\cite{gupta1997privatization,autoPrivatPeng,li_array_1992,feautrier_array_1988,midkiff_automatic_2012,maydan_array-data_1993,lefebvre_automatic_1998,Qui00,Darte_contraction_2005}.

\subsection{Background}

In this section, we provide an overview of two main concepts used in the polyhedral model: \emph{integer sets} and \emph{maps}. These two concepts will be used in later sections to define the different IR layers.

\emph{Integer sets} represent iteration domains while \emph{maps} are used to represent memory accesses and to transform iteration domains and memory accesses (apply loop nest and memory access transformations).  More details and formal definitions for these concepts are provided in~\cite{verdoolaege_isl:_2010,pencil_pact,polyhedral}.

An integer set is a set of integer tuples described using affine constraints.  An example of a set of integer tuples is \polyc{$\{(1,1); (2,1); (3,1); (1,2); (2,2); (3,2)\}$}
Instead of listing all the tuples as we do in the previous set, we can describe the set using affine constraints over loop iterators and symbolic constants as follows: \polyc{$\{S(i,j): 1 \leq i \leq 3 \wedge 1 \leq j \leq 2\}$} where $i$ and $j$ are the dimensions of the tuples in the set.

A map is a relation between two integer sets.  For example
\polyc{$\{S1(i,j) \rightarrow S2(i+2,j+2) : 1 \leq i \leq 3 \wedge 1 \leq j \leq 2\}$}
is a map between tuples in the set S1 and tuples in the set S2 (e.g. the tuple \poly{$S1(i,j)$} maps to the tuple \poly{$S2(i+2,j+2)$}).

All sets and maps in \framework are implemented using the Integer Set Library (ISL)~\cite{verdoolaege_isl:_2010}. We also use the ISL library notation for sets and maps throughout the paper.


\subsection{The Multi-Layer IR}

\begin{figure}
 \includegraphics[scale=0.41]{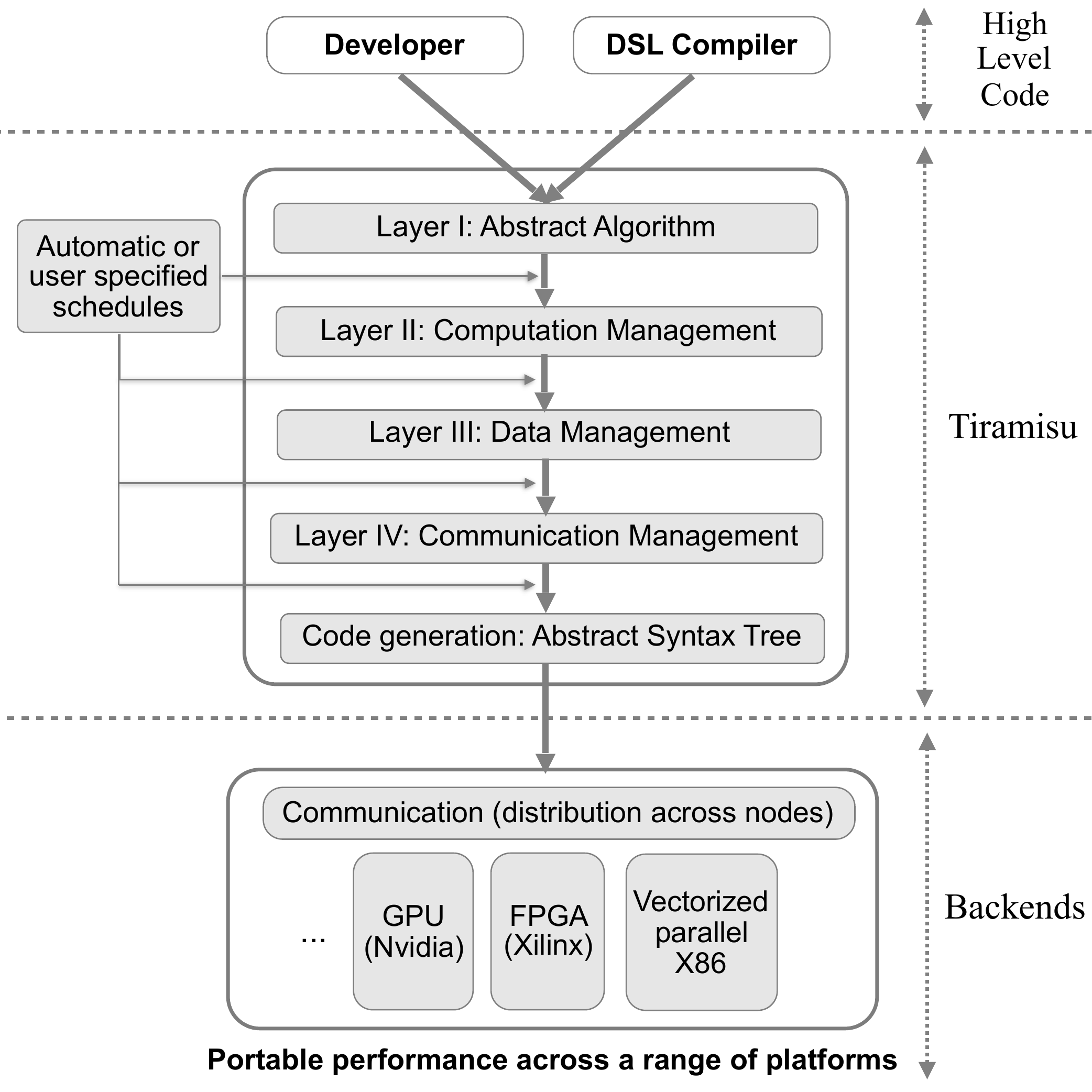}
 \caption{\framework overview}
 \label{fig:overview}
\end{figure}

A typical workflow for using \framework is illustrated in Figure~\ref{fig:overview}.  The user writes the pure algorithm and provides a set of scheduling commands.  The first layer of the IR is then transformed into lower layers, and finally \framework{} generates LLVM or other appropriate low-level IR.
\framework{} uses integer sets to represent each of the four IR layers and uses maps to represent transformations on the iteration domain and data layout.
The remainder of this section describes the four layers of the \framework IR.

\subsubsection{Layer I (\Layerone)}
\label{layer1}

Layer I of \framework{} specifies the algorithm without specifying when and where computations occur, how data should be stored in memory, or communication. Values are communicated via explicit producer-consumer relationships.

For example, the Layer I representation of the code in Figure~\ref{fig:algorithm} for the computation \texttt{by} is as follows:

\noindent \poly{$\{by(i, j, c): 0\leq i<N-2 \wedge 0\leq j<M-2 \wedge 0\leq c<3\} : (bx(i,j,c)+bx(i+1,j,c)+bx(i+2,j,c))/3$}

The first part, 
\poly{$\{by(i,j,c): 0\leq i<N-2 \wedge 0\leq j<M-2 \wedge 0\leq c<3\}$}, specifies the iteration domain of the computation \texttt{by}, while the second part 
is the computed expression.
The iteration domain is the set of tuples \poly{$by(i,j,c)$} such that \poly{$0\leq i< N-2 \wedge 0\leq j< M-2 \wedge 0\leq c < 3$}.
Computations in Layer I are not ordered;  declaration order does not affect the order of execution, which is specified in Layer II. 

\subsubsection{Layer II (\Layertwo)}
\label{layer2}

Layer II of \framework{} specifies the order of execution of computations and the processor on which they execute.  This layer does not specify how intermediate values are stored in memory; this simplifies optimization passes since these transformations do not need to perform complicated data-layout transformations.  The transformation of Layer I into Layer II is done automatically using scheduling commands.

Figure~\ref{fig:mainexample}-\codetwo{} (right) shows the GPU-optimized version of the code, produced by the scheduling and data-layout commands on the left side.
The corresponding Layer II representation for the \texttt{by} computation is shown below:

\noindent \poly{$\{ by(1, i0 (gpuB), j0 (gpuB), i1 (gpuT), j1 (gpuT), c) :  i0=floor(i/32) \wedge j0=floor(j/32) \wedge i1=i\%32 \wedge j1=j\%32 \wedge 0\leq i<N-2 \wedge 0\leq j<M-2 \wedge 0\leq c<3\}: (bx(i0*32+i1,j0*32+j1,c)+bx(i0*32+i1+1,j0*32+j1,c)+bx(i0*32+i1+2,j0*32+j1,c))/3$}

Computations in Layer II are ordered based on their lexicographical order\footnote{For example the computation $S0(0, 0, 0)$ is lexicographically before the computation \mbox{$S0(0, 0, 1)$} and the computations $S0(0, i, 0)$ are lexicographically before the computations $S0(1, i, 0)$}.  The set before the colon in the representation is an ordered set of computations.
The tag \emph{gpuB} for the dimension $i0$ and $j0$ indicates that each iteration ($i0,j0$) is mapped to the GPU block ($i0,j0)$. In Layer II, the total ordering of these tuples determines execution order.

Computations in this layer are ordered and assigned to a particular processor; the order is dictated by \textit{time dimensions} and \textit{\processor dimensions}.  Time dimensions specify the order of execution relative to other computations while \processor{} dimensions specify on which processor each computation executes.
\Processor{} dimensions are distinguished from time dimensions using tags, which consist of a processor type.  Currently, \framework{} supports the following space tags:

{
\centering
{
    \footnotesize
    \setlength\tabcolsep{5pt}
    \vspace{0.5cm}
    \begin{tabular}{ll}
        \texttt{cpu} & the dimension runs on a CPU in a shared memory system \\
        \texttt{node} & the dimension maps to nodes in a distributed system \\
        \texttt{gpuT} & the dimension maps to a gpu thread dimension. \\
        \texttt{gpuB} & the dimension maps to a gpu block dimension.\\
    \end{tabular}
    \vspace{0.5cm}
}
}

Tagging a dimension with a processor type indicates that the dimension will be distributed over processors of that type; for example, tagging a dimension with \emph{cpu} will execute each iteration of that loop dimension on a separate CPU.

Other tags that transform a dimension include:

{
\centering
{
    \footnotesize
    \setlength\tabcolsep{5pt}
    \vspace{0.5cm}
    \begin{tabular}{ll}
        \texttt{vec(s)} & vectorize the dimension (\emph{s} is the vector length)\\
        \texttt{unroll} & unroll the dimension\\
    \end{tabular}
    \vspace{0.5cm}
}
}

Computations mapped to the same processor are ordered by projecting the computation set onto the time dimensions and comparing their lexicographical order.

\subsubsection{Layer III (\Layerthree)}
\label{layer3}

Layer III makes the data layout concrete by specifying where intermediate values are stored.  Any necessary buffer allocations/deallocations are also constructed in this level.  \framework{} generates this layer automatically from Layer II by applying the scheduling commands for data mapping.

The \layerthree layer specifies memory locations for storing computed values.  It consists of the Layer II representation along with allocation/deallocation statements, and a set of \emph{access relations},
which map a computation from Layer II to array elements read or written by that computation.  Scalars are treated as single-element arrays.  
For each buffer, an allocation statement is created, specifying the type of the buffer and its size.  Similarly, a deallocation statement is also added.

Possible data mappings in \framework include mapping computations to structures-of-arrays, arrays-of-structures, and contraction of multidimensional arrays into arrays with fewer dimensions or into scalars.  It is also possible to specify more complicated accesses such as the storage of computations $c(i,j)$ into the array elements $c(i\%2,j\%2)$ or into $c(j,i)$.


In the example of Figure~\ref{fig:mainexample}-\codetwo{} (left), setting the data access using \texttt{by.store\_in({c,i,j})}
indicates that the result of the computation \poly{$by(i, j, c)$} is stored in the array element \poly{$by[c,i,j]$}. This command generates the following map in Layer III:

\noindent
\poly{$\{by(1, i0 (gpuB), j0 (gpuB), i1 (gpuT), j1 (gpuT), c)\rightarrow by[c,i0*32+i1,j0*32+j1]: i0=floor(i/32) \wedge j0=floor(j/32) \wedge i1=i\%32 \wedge j1=j\%32 \wedge 0\leq i<N-2 \wedge 0\leq j<M-2 \wedge 0\leq c<3\}$}

Data mapping in \framework is an affine relation that maps each computation to a buffer element.  \framework allows any data-layout mapping expressible as an affine relation.

\subsubsection{Layer IV (\Layerfour)}
\label{layer4}

Layer IV adds synchronization and communication operations to the representation, mapping them to the time-\processor domain,  and concretizes when statements for buffer allocation/deallocation occur.  This layer is generated automatically from Layer III by applying user-specified commands.
Any allocation or deallocation operation added in Layer III is also mapped to the time-\processor domain in this layer.

\section{Compiler Implementation}

Since the main contribution of this paper is not in introducing new techniques for code generation, we only provide a high level overview of how \framework{} generates the IR layers and target code.  Throughout the section, we refer the reader to the appropriate literature for more details.



In the rest of this section we describe how scheduling commands transform Layers I, II, III and IV.   We also describe how target code is generated from Layer IV.

\paragraph{Transforming Layer I into Layer II}
Transforming Layer I into Layer II is done using two types of scheduling commands: (1) commands for loop nest transformations (such as \texttt{tile()}, \texttt{split()}, \texttt{shift()}, \texttt{interchange()}); and (2) commands for mapping loop levels to hardware (including\\ \texttt{parallelize()}, \texttt{vectorize()}, \texttt{gpu()}).

The first type of scheduling command applies a map that transforms the iteration domain.  For example, when a tiling command is applied on the \texttt{by} computation in Figure~\ref{fig:algorithm}, it gets translated into the following map:

\centerline{\poly{$\{by(i,j,c) \rightarrow by(i0,j0,i1,j1,c): i0=floor(i/32) \wedge i1=i\%32 \wedge $}}
\centerline{\polyc{$ j0=floor(j/32) \wedge j1=j\%32 \wedge 0 \leq i<N \wedge 0 \leq j<N\}$}}

This map is then applied on the Layer I representation, producing the Layer II representation. Composing transformations is done by composing different maps, since the composition of two affine maps is an affine map.

The second type of command adds \processor tags to dimensions to indicate which loop levels to parallelize, vectorize, map to GPU blocks, and so on.

\paragraph{Transforming Layer II into Layer III}
This is done by augmenting Layer II with access relations.  By default, \framework{} uses identity access relations (i.e., access relations that store a computation \texttt{C(i,j)} into a buffer \texttt{C[i,j]}).  If the \texttt{store\_in()} command is used, the access relation is deduced from that command instead.  Buffer allocations are also added while transforming Layer II into Layer III. The scheduling command \texttt{b.allocate\_at(C, i)} creates a new statement that allocates the buffer \texttt{b} in the same loop nest of the computation \texttt{C} but at loop level \texttt{i}.

\paragraph{Transforming Layer III into Layer IV}
Scheduling commands for data communication (send and receive), synchronization, and for copying data between global, shared and local memory are all translated into statements.  For example, the \texttt{send()} and \texttt{receive()} commands are translated into function calls that will be translated into MPI calls during code generation.

\subsection{Code Generation}

Generating code from the set of computations in Layer IV amounts to generating nested loops that visit each computation in the set, once and only once, while following the lexicographical ordering between the computations~\cite{Bas04,Iri88,Qui00}. \framework{} relies on an implementation of the Cloog~\cite{Bas04} code generation algorithm provided by the ISL library~\cite{verdoolaege_isl:_2010}. 
The \framework{} code generator takes Layer IV IR and generates an abstract syntax tree (AST).  The AST is then traversed to generate lower level code for specific hardware architectures (depending on the target backend).


The multicore CPU code generator generates LLVM IR from the AST. In order to generate LLVM IR, we use Halide as a library: we first generate the Halide IR then we lower the Halide IR to LLVM IR using Halide. We do not use Halide to perform any high level code optimization. All the code optimizations are performed by \framework{} before generating the Halide IR. The Halide compiler then lowers the Halide IR loops into LLVM IR.


The GPU code generator generates LLVM IR for the host code  and CUDA for the kernel code. Data copy commands and information about where to store buffers (shared, constant, or global memory) are all provided in Layer IV.  \framework{} translates these into the equivalent CUDA data copy calls and buffer allocations in the generated code.  Computation dimensions tagged with GPU thread or GPU block tags are translated into the appropriate GPU thread and block IDs in the lowered code.  The \framework{} code generator can generate coalesced array accesses and can use shared and constant memories. It can also avoid thread divergence by separating full tiles (loop nests with a size that is multiple of the tile size) from partial tiles (the remaining part of a loop).


The code generator for distributed memory systems utilizes MPI.  During code generation, all the function calls for data copying are translated to the equivalent MPI function calls.  The generated code is postprocessed and each distributed loop is converted into a conditional based on the MPI rank of the executing process. For example:

\begin{lstlisting}[numbers=none]
for(q in 1..N-1) {...} // distribute on q
\end{lstlisting}

becomes:

\begin{lstlisting}[escapechar=@,numbers=none]
q = get_rank(); if (q@$\geq$1@ and q<N-1) {...}
\end{lstlisting}



\subsection{Support for Non-Affine Iteration Spaces\label{nonaffine}}

\framework represents non-affine array accesses, non-affine loop bounds, and non-affine conditionals in a way similar to Benabderrahmane et al.~\cite{Benabderrahmane}.
For example, a conditional is transformed into a predicate and attached to the computation.  The list of accesses of the computation is the union of the accesses of the computation in the two branches of the conditional; this is an over-approximation. During code generation, a preprocessing step inserts the conditional back into the generated code.  The efficiency of these techniques was demonstrated by Benabderrahmane et al.~\cite{Benabderrahmane} and was confirmed in the PENCIL compiler~\cite{pencil}. Our experiences in general, as well as the experiments in this paper, show that these approximations do not hamper performance.

\section{Evaluation}
\label{sec:eval}

We evaluate \framework{} on two sets of benchmarks. The first is a set of deep learning and linear algebra benchmarks.
The second is a set of image processing benchmarks.

We performed the evaluation on a cluster of 16 nodes. Each node is a dual-socket machine with two 24-core Intel Xeon E5-2680v3 CPUs, 128 GB RAM, Ubuntu 14.04, and an Infiniband interconnect.  We use the MVAPICH2 2.0 \cite{mvapich2} implementation of MPI for the distributed tests.
The multicore experiments (CPU) are performed on one of these nodes.
GPU experiments are performed on an NVIDIA Tesla K40 with 12 GB of RAM.  Each experiment is repeated $30\times$ and the median time is reported.

\subsection{Deep Learning and Linear Algebra Benchmarks}

\begin{figure}[t]
\centering
\includegraphics[width=1\columnwidth]{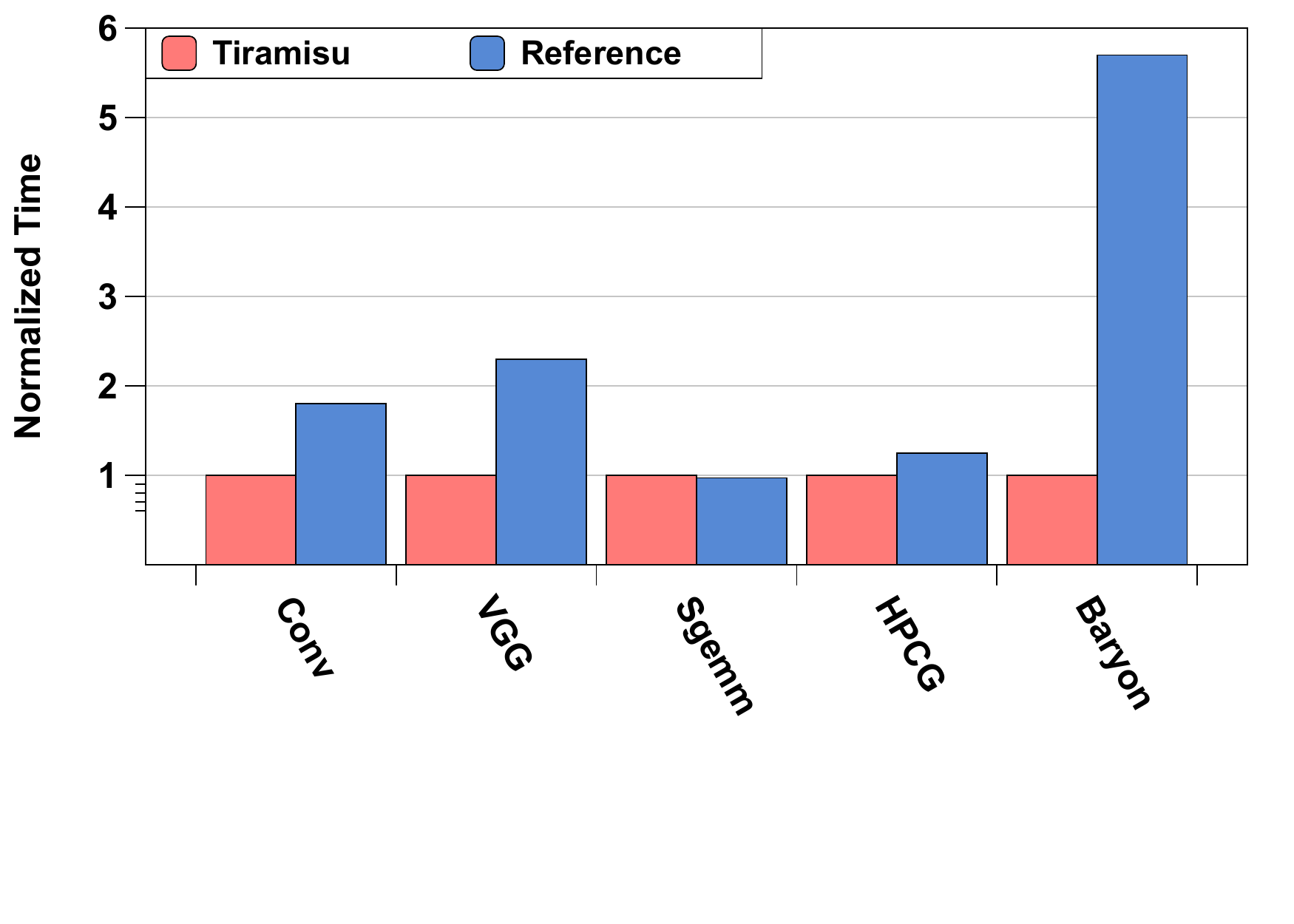}
\vspace{-1.7cm}
\caption{Normalized Execution Times for Deep Learning, Linear and Tensor Algebra Benchmarks.}
\label{fig:mkl}
\end{figure}

We evaluated \framework{} by implementing a set of deep learning and linear algebra benchmarks, including \texttt{Conv} (a direct implementation of a neural network convolution layer), \texttt{VGG} (a block of a VGG neural network), and \texttt{sgemm} (matrix multiplication used to implement convolutions), \texttt{HPCG} (a benchmark for multigrid preconditioned conjugate gradient, CG)\footnote{\url{http://www.hpcg-benchmark.org/}}, and \texttt{Baryon} (a dense tensor contraction code for constructing Baryon Building Blocks~\cite{detmold2013nuclear}).
For all of these benchmarks, we compare the \framework{} implementation with Intel MKL, except for \texttt{HPCG} and \texttt{Baryon}, where we compare \framework{} with reference implementations.
Figure~\ref{fig:mkl} shows a comparison between the performance of CPU code generated by Tiramisu and reference code. For \texttt{sgemm} and \texttt{HPCG} we use matrices of size $1060\times1060$ and vectors of size 1060 while for \texttt{Conv} and \texttt{VGG} we use $512\times512$ as the data input size, 16 as the number of input/output features and a batch size of 32. For \texttt{Baryon}, we use the same tensor sizes as in the reference code.

For \texttt{sgemm}, \framework{} matches the performance of Intel MKL. \texttt{sgemm} is interesting in particular because the Intel MKL implementation of this kernel is well-known for its hand-optimized performance.
We used a large set of optimizations to match Intel MKL.
These optimizations include two-level blocking of the three-dimensional \texttt{sgemm} loop,
vectorization, unrolling, array packing, register blocking, and separation of full and partial tiles (which is crucial to enable vectorization, unrolling, and reducing control overhead). We also used auto-tuning to find the best tile size and unrolling
factor for the machine on which we run our experiments.

For the \texttt{Conv} kernel, \framework{} outperforms the
Intel MKL implementation because the \framework{}-generated code uses a fixed size for the convolution filter.  We generate  specialized versions for common convolution filter sizes ($3\times3$, $5\times5$, $7\times7$, $9\times9$ and $11\times11$). This allows the \framework{} compiler to apply optimizations that Intel MKL does not perform; for example this allows \framework{} to unroll the innermost (convolution filter) loops since their size is known at compile time. In \texttt{VGG}, \framework{} fuses the two convolution loops of the \texttt{VGG} block, which improves data locality.  In addition, we generate code with fixed sizes for convolution filters (as we did in \texttt{Conv}). This provides $2.3\times$ speedup over Intel MKL.
The \framework{} speedup over the \texttt{Baryon} reference code is achieved through vectorization, but this vectorization is not trivial since it requires the application of array expansion and then the use of scatter/gather operations, which are both not implemented in the reference \texttt{Baryon} code.

\subsection{Image Processing Benchmarks}

We used the following image processing benchmarks in our evaluation: \texttt{edgeDetector}, a ring blur followed by Roberts edge detection~\cite{roberts65}; \texttt{cvtColor}, which converts an RGB image to grayscale; \texttt{conv2D}, a simple 2D convolution; \texttt{warpAffine}, which does affine warping on an image;  \texttt{gaussian}, which performs a gaussian blur; \texttt{nb}, a synthetic pipeline composed of 4 stages that computes a negative and a brightened image from the same input image; and \texttt{ticket \#2373}, a code snippet from a bug filed against Halide. This code simply has a loop that assigns a value to an array but the iteration space is not rectangular (it tests if \texttt{x >= r} where \texttt{x} and \texttt{r} are loop iterators).  The inferred bounds in this code are over-approximated, causing the generated code to fail due to an assertion during execution.
Four of these benchmarks have non-affine array accesses and non-affine conditionals for clamping (to handle boundary cases): \texttt{edgeDetector}, \texttt{conv2D}, \texttt{warpAffine} and \texttt{gaussian}.
We used a $2112\times3520$ RGB input image for the experiments.

\begin{figure*}[h!t]
\centering
\includegraphics[width=1.3\columnwidth,trim=50 10 10 10]{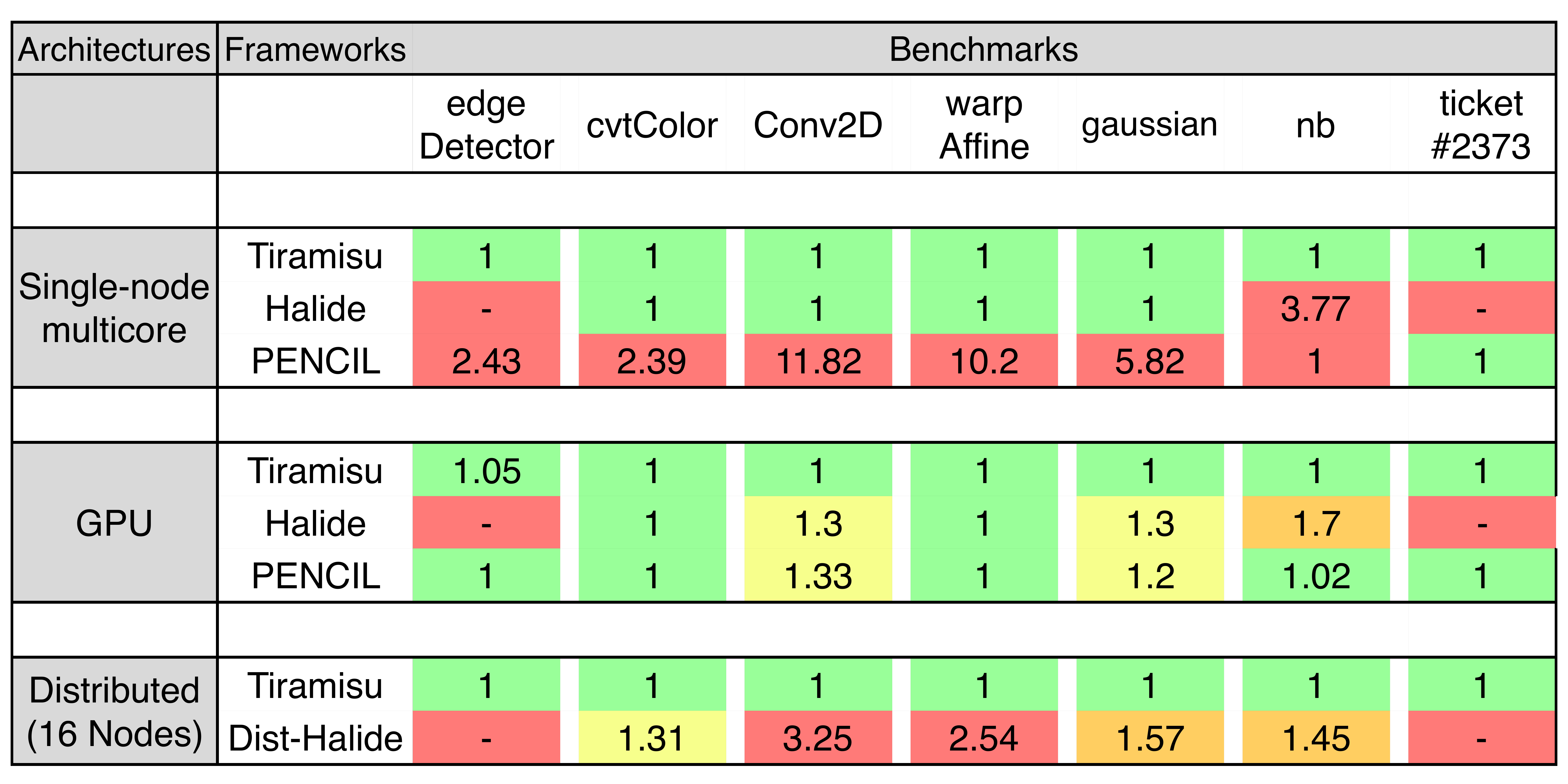}
\caption{A heatmap comparing the normalized execution times of code generated by \framework{} with other frameworks (lower is better).  Comparison is performed on three architectures: single-node multicore, GPU, distributed (16 nodes). "-" indicates unsupported benchmarks.}
\label{fig:speedup}
\end{figure*}

We compare \framework{} with two other compilers: Halide~\cite{halide_12},  an industrial-quality DSL for image processing that has a scheduling language, and PENCIL~\cite{pencil_paper}, a state-of-the-art fully automatic polyhedral compiler.



Figure~\ref{fig:speedup} compares the normalized execution time of code generated by \framework{} to other state-of-the-art frameworks on three architectures: single-node multicore, GPU and distributed (16 nodes).  For the single-node multicore and GPU we compare \framework{} to Halide and PENCIL.  For the distributed architecture, we compare to distributed Halide~\cite{denniston2016distributed}.

\paragraph{Single-node multicore}
In four of the benchmarks, the performance of the code generated by \framework{} matches the performance of Halide.  We use the same schedule for both implementations; these schedules were hand-written by Halide experts.  The results for \texttt{edgeDetector}, \texttt{conv2D}, \texttt{warpAffine} and \texttt{gaussian}, which have non-affine array accesses and conditionals, show that \framework{} handles such access patterns efficiently.

Two of the other benchmarks, \texttt{edgeDetector} and \texttt{ticket \#2373}, cannot be implemented in Halide.  The following code snippet shows \texttt{edgeDetector}:

\vspace{0.25cm}
\begin{lstlisting}[language=C,escapechar=@,numbers=none]
/* Ring Blur Filter */
R(i,j) =(Img(i-1,j-1) + Img(i-1,j) + Img(i-1,j+1)+
         Img(i,j-1)   +              Img(i,j+1)  +
         Img(i+1,j-1) + Img(i+1,j) + Img(i+1,j+1))/8
/* Roberts Edge Detection Filter */
Img(i,j) = abs(R(i,j)  - R(i+1,j-1)) +
           abs(R(i+1,j)- R(i,j-1))
\end{lstlisting}
\vspace{0.25cm}

\texttt{edgeDetector} creates a cyclic dependence graph with a cycle length $\geq 1$ ( \texttt{R} is written in the first statement and read in the second while \texttt{Img} is written in the second and read in the first), but
Halide can only express programs with an acyclic dependence graph, with some exceptions;  this restriction is imposed by the Halide language and compiler to avoid the need to prove the legality of some optimizations (since proving the legality of certain optimizations is difficult in the Halide interval-based representation).
\framework{} does not have this restriction since it checks transformation legality using dependence analysis~\cite{feautrier_dataflow_1991}.

In \texttt{ticket \#2373}, which exhibits a triangular iteration domain,  Halide's bounds inference over-approximates the computed bounds, which leads the generated code to fail in execution.  This over-approximation in Halide is due to the use of intervals to represent iteration domains, which prevents Halide from performing precise bounds inference for non-rectangular iteration spaces.  \framework{} can handle this case naturally since it relies on the polyhedral model where sets can include any affine constraint  in addition to loop bounds.  These examples show that the model exposed by \framework{} naturally supports more complicated code patterns than an advanced, mature DSL compiler.

For \texttt{nb}, the code generated from \framework{} achieves $3.77\times$ speedup over the Halide-generated code. This is primarily due to loop fusion.  In this code, \framework{} enhances data locality by fusing loops into one loop;  this is not possible in Halide, which cannot fuse loops if they update the same buffer.  Halide makes this conservative assumption because otherwise it cannot prove the fusion is legal. This is not the case for \framework{}, which uses dependence analysis to prove correctness.

The slowdown of the PENCIL compiler in \texttt{gaussian} is due to a suboptimal decision made by PENCIL.  The \texttt{gaussian} kernel is composed of two successive loop nests (each of them contains three loop levels). PENCIL decides to interchange the two innermost loop levels in order to enable the fusion of the two successive loop nests. This decision minimizes the distance between producer and consumer statements (first and second loop nests), but it also reduces spatial locality because it leads to non-contiguous memory accesses.  The right decision in this case is a trade-off.  Such a trade-off is not captured by the Pluto automatic scheduling algorithm used within PENCIL.
For the other kernels, both \framework{} and Halide apply vectorization and unrolling on the innermost loops, while PENCIL does not since the multicore code generator of PENCIL does not implement these two optimizations.  For \texttt{warpAffine}, both \framework{} and Halide have a high speedup over PENCIL because the unique loop nest in this benchmark has 25 statements, and vectorizing the innermost loop transforms all of these statements to their vector equivalent while unrolling increases register reuse and instruction level parallelism on the $24$ cores of the test machine.

\paragraph{GPU}
For the GPU backend, the reported times are the total execution times (data copy and kernel execution).
Code generated by \framework{} for \texttt{conv2D} and \texttt{gaussian} is faster than that of Halide because code generated by \framework{} uses constant memory to store the weights array, while the current version of Halide does not use constant memory for its PTX backend.  The only difference between the schedule of \framework{} and Halide in these benchmarks is the use of \texttt{tag\_gpu\_constant()} in \framework{}.  Data copy times, for all the filters,  are the same for \framework{} and Halide.  For \texttt{nb}, the code generated by \framework{} achieves $1.7\times$ speedup over that generated by Halide because \framework{} is able to apply loop fusion, which Halide cannot apply.

Compared to PENCIL, the speedup in \texttt{conv2D} and \texttt{gaussian} is due to the fact that PENCIL generates unnecessarily complicated control flow within the CUDA kernel, which leads to thread divergence.

\paragraph{Distributed}
We assume the data are already distributed across the nodes by rows. Of these benchmarks, \texttt{nb}, \texttt{cvtColor} and \texttt{ticket \#2373} do not require any communication; the other four require communication due to overlapping boundary regions in the distributed data.

Figure \ref{fig:speedup} compares the execution time of distributed \framework{} and distributed Halide. \framework{} is faster than distributed Halide in each case.  It achieves up to $3.25\times$ speedup for \texttt{conv2D}.  For the kernels involving communication, code generated by distributed Halide has two problems compared to \framework{}:  distributed Halide overestimates the amount of data it needs to send, and unnecessarily packs together contiguous data into a separate buffer before sending.  

\begin{figure}[t]
\centering
\includegraphics[width=0.9\columnwidth, trim=0 30 0 60]{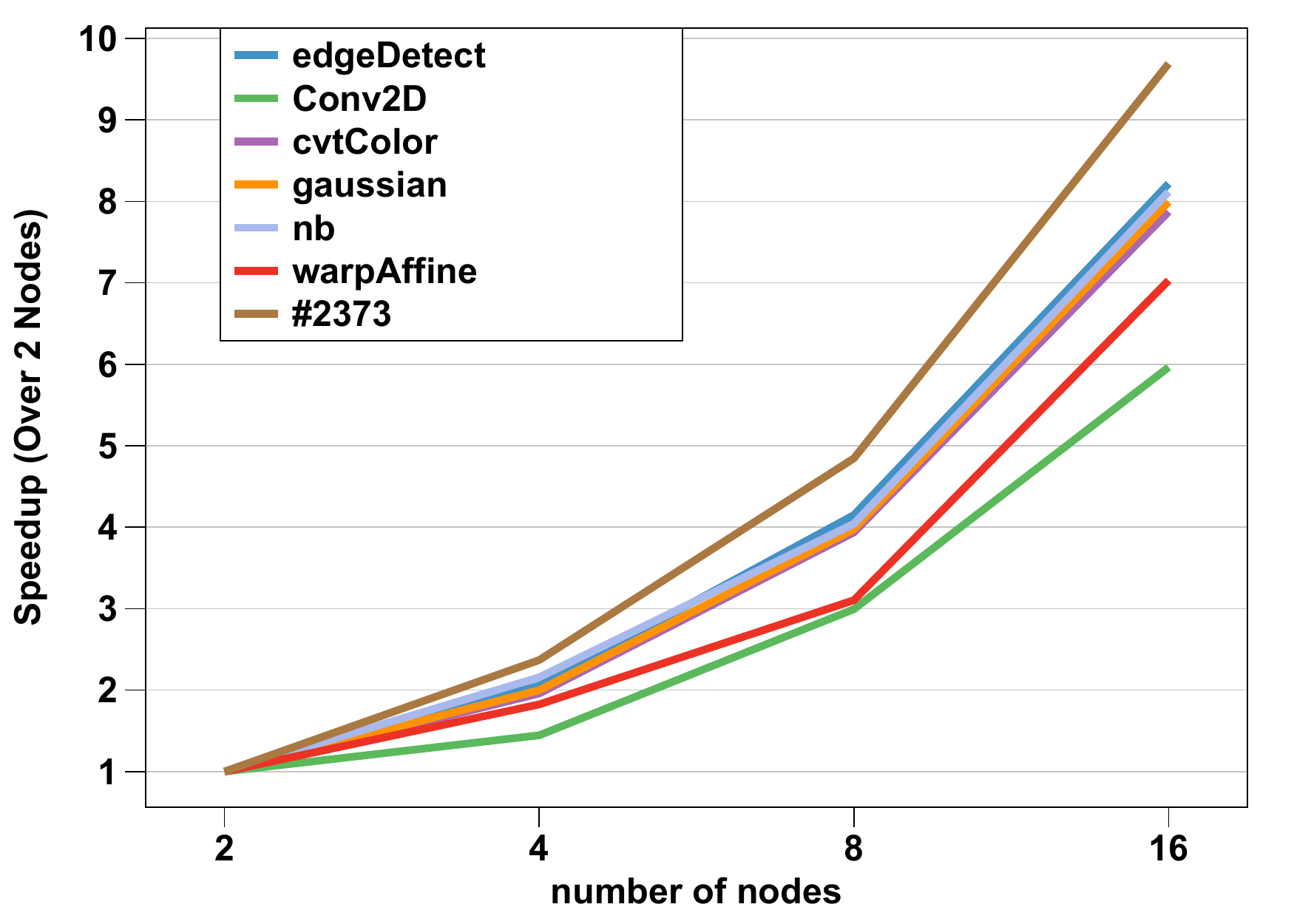}
\vspace{0.4cm}
\caption{Speedup of code generated by distributed \framework{} for 2, 4, 8, and 16 nodes. The baseline is the execution time on 2 nodes.}
\label{fig:distCPU_tiramisu_scaling_exec}
\end{figure}


Distributed Halide overestimates the amount of data it needs to send because the benchmarks have array accesses that cannot be analyzed statically (the array accesses are clamped\footnote{\texttt{clamp(i, 0, N)} returns $0$ if $i<0$, $N$ if $i>N$, $i$ otherwise.} to handle boundary cases), therefore distributed Halide cannot compute the exact amount of data to send.
To avoid this problem, \framework{} uses explicit communication using the \texttt{send()} and \texttt{receive()} scheduling commands.  The use of these two commands is the only difference between the \framework{} and distributed Halide.  These commands allow the user to specify exactly the amount of data to send and also allow the compiler to avoid unnecessary packing.

Figure \ref{fig:distCPU_tiramisu_scaling_exec} shows the speedup of the kernels with distributed \framework{} when running on 2, 4, 8, and 16 nodes.  This graph shows that distributed code generated from \framework{} scales well as the number of nodes increases (strong scaling).



\section{Conclusion}

This paper introduces \framework, a polyhedral compiler framework that features a scheduling language with commands for targeting multicore CPUs, GPUs, and distributed systems.  A four-layer intermediate representation that separates the algorithm, when and where computations occur, the data layout and the communication is used to implement the compiler.
We evaluate \framework by targeting a variety of backends and demonstrate that it generates code matching or outperforming state-of-the-art frameworks and hand-tuned code.

\newpage

\section*{Acknowledgements}
This work was supported by the ADA Research Center, a JUMP Center co-sponsored by SRC and DARPA.

\bibliography{biblio}

\end{document}